\documentclass[prl,twocolumn]{revtex4}

\usepackage{amsmath}
\usepackage{epsfig}

\date{\today}

\begin{document}

\begin{abstract}
  We present a new method for measuring the CMB temperature quadrupole, using
  large scale CMB polarization. The method exploits the fact that CMB
  polarization is partially sourced by the local temperature quadrupole. We link the
  temperature with the polarization spectrum directly by relating the local
  quadrupole at the onset of reionization to both of them. The dominant
  contribution is at $l<30$ and since we use many $l$ values, we can reduce
  the error significantly below cosmic variance. In particular, for our 
  fiducial model, the error on the temperature quadrupole is reduced to $24\%$. This 
  has the potential of reducing the probability of a low quadrupole by two orders
  of magnitude.
\end{abstract}

\title{A new method for measuring the CMB temperature quadrupole with an
accuracy better than cosmic variance} 

\author{Constantinos Skordis}
\email{skordis@astro.ox.ac.uk}

\author{Joseph Silk} \email{silk@astro.ox.ac.uk}

\affiliation{University of Oxford}

\maketitle

How precisely can $C^T_2$, the {\it Cosmic Microwave Background} (CMB)
temperature quadrupole, be measured? At first sight, the question seems
trivial to answer~: the measurement is limited by cosmic variance, which at
the quadrupole is equal to $\sqrt{\frac{2}{5}} C^T_2$.  The first release of  {\it
Wilkinson Microwave Anisotropy Probe} (WMAP) data~\cite{WMAP} has
regenerated interest in the value of the CMB quadrupole, as WMAP measured a
lower quadrupole than expected, based on a $\Lambda$CDM cosmology.  This
confirmed the {\it COsmic Background Explorer} (COBE)~\cite{COBE} measurement
but more cleanly, as the lower detector noise and wider frequency
range to pin down galactic foreground emission~\cite{WMAPfor} renders the
measurement more robust.

Let us consider the value of the quadrupole $\Delta T_2^2$ where $\Delta T_l^2
 = \frac{l(l+1)}{2\pi} C_l$. Consider also the best fit adiabatic model
 of~\cite{BDFMS} as a fiducial model for the rest of the paper, a six
 parameter model with physical baryon and cold dark matter densities of
 $\omega_b = 0.023$ and $\omega_c=0.117$ respectively, relative cosmological
 constant density $\Omega_\Lambda=0.715$, optical depth to reionization
 $\tau=0.137$ and scalar spectral index $n = 0.974$. Using frequentist
 statistics given the fiducial model and assuming full sky coverage with
 cosmic variance as the only source of error, the probability of measuring a
 quadrupole as low as or lower than the quoted WMAP value of $\Delta T_2^2 = 123.4 \;(\mu
 K)^2$ is $0.01$. Using a sky cut would actually raise the
 probability. Efstathiou~\cite{Efstathiou1} has argued that statements such
 as the one above should not be taken too seriously and do not disfavor the
 $\Lambda$CDM concordance model.  If for example, one uses $\Delta T_2^2 =
 201.6 \;(\mu K)^2$ quoted by Tegmark {\it et al.}~\cite{TdOCH} using their
 different method of foreground subtraction, the probability rises to
 $0.03$. Efstathiou also showed that the value of the quadrupole is sensitive
 to the estimator used~\cite{Efstathiou2}.  In particular, the use of the
 quadratic estimator~\cite{QML} rather than the pseudo-$C_l$ estimator used
 by WMAP~\cite{PCL} is better suited for low $l$ values and could further
 increase the quadrupole to $\Delta T_2^2 = 250 \;(\mu K)^2$. The
 corresponding probability for the fiducial model would then increase to
 $0.05$, {\it five times larger} than the originally quoted value.  Other
 apparent coincidences such as  the alignment between the quadrupole and the
 octopole~\cite{dOCTZH}, the north-south asymmetry~\cite{EHBGL} or possible
 non-Gaussianity/global anisotropy~\cite{CHS} do not concern us here even
 though a {\it normal} quadrupole would certainly weaken these findings.

The probability arguments discussed above depend strongly on the cosmic
variance limit. If there were a way to measure $C_2^T$ with precision better
than cosmic variance, the underlying probability distribution of the
quadrupole would be different and the probabilities could change drastically.
Is there a different measurement one might  make, that could measure the CMB
temperature quadrupole and has an error smaller than cosmic variance? This
question has been raised and partially answered by Kamionkowski and
Loeb~\cite{KL}. They considered the polarization spectrum produced by
clusters through the Sunyaev-Zel'dovich effect~\cite{SZ}.  Since the
polarization spectrum depends on the local quadrupole at the cluster, one can
get information on the quadrupole $C_2^T(z)$ at redshift $z$, by taking large samples of
clusters and taking an average. One hopes that the averaging method will
recover a quadrupole close to the true cosmological value.  The authors
mention another way of getting a measure of $C_2^T$, namely large angle CMB
polarization, generated by reionization. They argue however that this would
depend on the reionization details and will therefore be very difficult to
handle.  Further calculations involving clusters have been carried out
in~\cite{SS,CB} and more recently in~\cite{P}.

The connection between the temperature quadrupole and polarization has been
exploited further by Dor\'e et. al.~\cite{DHL}. They considered the  consistency
of the observed temperature spectrum with the polarization-temperature cross
correlation, given the best fit model of WMAP. In particular, if the observed
temperature quadrupole was anomalously low, then one would expect low large angle
polarization power as well. Their method consists of Monte-Carlo realizations
of their fiducial model combined with the well known statistical correlations
between the two above spectra~\cite{ZSS}. Using a frequentist approach, they
find that the two are inconsistent at the $98.5\%$ level. This hints that the
low temperature quadrupole might not be just due to an unlucky throw of dice.

In this paper, we exploit the same connection between temperature and
polarization as above and demonstrate, using a different method, the
possibility of getting a measure on $C_2^T$ from large angle polarization
generated by reionization. We show that contrary to the argument
in~\cite{KL}, the method does not depend on the details of reionization but
rather on the initial amplitude of the local quadrupole at the onset of
reionization. Any reionization history dependence can be accommodated by
using extra parameters but this does not appear to be an obstacle to our
method. That the details of reionization can be observed in the CMB has been
studied before~\cite{reion}.  Even in the worst case scenario, the
information about reionization contained in the CMB boils down to at most
five parameters~\cite{HH} but this is something that would have to be taken
into account for standard CMB parameter extraction as well. Therefore in what
follows, we use a sharp reionization transition at (conformal) time $t_r$,
the time where the electron ionization fraction first rises above its
residual recombination value.

We first  review how the relevant spectra are related.  Let
$\Delta^T_l(k,t)$ and $\Delta^P_l(k,t)$ denote the $l$-multipole of the
temperature and polarization transfer functions respectively. Let also $t$ be
conformal time with $t_0$ being the time today, $k$  the wavenumber and
$\tau(t) = \int^t_{t_0} \dot{\tau} \; dt$  the optical depth to time $t$
where a dot indicates differentiation with respect to conformal time.

The temperature quadrupole is then given in terms of an initial power
spectrum $P_\Psi(k)$ and transfer function $\Delta^T_2(k)$ as
\begin{equation}
C_2^T = \frac{2}{\pi}\int_0^\infty dk \; k^2 \; P_\Psi(k) \;
|\Delta^T_2(k)|^2 \;.
\end{equation}
The quadrupole's transfer function obeys the differential equation
\begin{equation}
  \dot{\Delta}^T_2 + \frac{9}{10}\dot{\tau}\Delta^T_2 =
      \frac{k}{5}\left(2\Delta^T_1 - 3\Delta^T_3 \right) +
      \frac{\dot{\tau}}{10}\left( \Delta^P_0 + \Delta^P_2 \right)\;.
      \label{eq:delta_2}
\end{equation}
The general solution is the sum of an initial condition (the initial
quadrupole) multiplied by the relevant Green's function and a particular
integral $\Delta_2^{PI}(k)$ over the source $S_2$ which is
given by the RHS of (\ref{eq:delta_2}). For reasons to become
clearer below, we choose the initial time to be the reionization time $t_r$,
so that the initial condition is the local quadrupole at the onset of
reionization. The local quadrupole today is then given by
\begin{equation}
  \Delta^T_2(k,t_0) = e^{\frac{9}{10}\tau_r} \Delta^T_2(k,t_r) +
  \int_{t_r}^{t_0} \; dt \; e^{\frac{9}{10}\tau(t)} \; S_2(k,t).
  \label{eq:delta_2_sol}
\end{equation}

\begin{figure}
 \epsfig{file=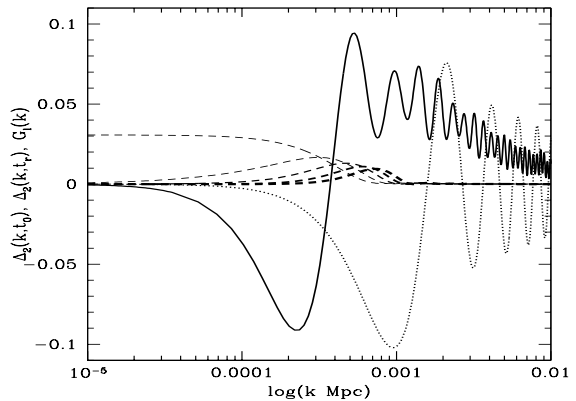,height=2.2in,width=3in}
\caption{The local quadrupole today(solid) and at reionization(dotted). Also shown
 are the first five $G_l$'s ($l=2$ to $l=6$) in dashed with increasing strength.}
\label{fig:quads}
\end{figure}

Similarly the E-mode polarization spectrum is given in terms of the same
initial power spectrum and E-mode transfer function $E_l(k)$ as
\begin{equation}
C_l^E = \frac{2}{\pi}\int_0^\infty dk \; k^2 \; P_\Psi(k) \; |E_l(k)|^2 \;.
\end{equation}
The E-mode transfer function is given by the line-of-sight integral along the
past light cone as~\cite{ZS}
\begin{equation}
E_l(k) = \frac{3}{4}\sqrt{\frac{(l+2)!}{(l-2)!}}\int_0^{t_0} dt \;
   \frac{j_l(x)}{x^2} \; \dot{\tau}e^{\tau} \; \Pi(k,t) \; , \label{eq:Emode}
\end{equation}
 where $x\equiv k(t_0 - t)$, $\Pi(k,t)$ is the polarization source and
$j_l(x)$ is the spherical Bessel function.  The polarization source is given
in terms of temperature quadrupole and polarization monopole and quadrupole
as
\begin{equation}
  \Pi(k,t) = \Delta_2^T(k,t) + \Delta^P_0(k,t) + \Delta^P_2(k,t)
	\label{eq:Polterm}
\end{equation}
and obeys the differential equation
\begin{equation}
  \dot{\Pi} + \frac{3}{10}\dot{\tau}\Pi = \frac{k}{5} \left[ 2\Delta^T_1 -
3\left( \Delta^T_3 + \Delta^P_1 + \Delta^P_3\right) \right] \;.
	\label{eq:Pol_diff}
\end{equation}

The general solution of the above equation for any time $t>t_r$ is
\begin{eqnarray}
  \Pi(k,t) &=& \;e^{\frac{3}{10}[\tau_r - \tau(t)]} \Delta^T_2(k,t_r)
  \nonumber \\ &&+ e^{-\frac{3}{10}[\tau(t)]} \int_{t_r}^{t} \; dt^{\prime}
  \; e^{\frac{3}{10}\tau(t^{\prime})} \; S_\Pi(k,t^{\prime}) \; .
  \label{eq:Polterm_sol}
\end{eqnarray}
where again we have chosen the initial time to be $t_r$ and where $S_\Pi$ is
given by the RHS of (\ref{eq:Pol_diff}).  Note that in the
above equation we have replaced $\Pi(k,t_r)$ which should have been the true
initial condition with $\Delta^T_2(k,t_r)$. It turns out that due to
free-streaming from recombination to reionization, the two are equal to one part
in $10^6$. During reionization, however, this is no longer true as the three
terms comprising $\Pi(k,t)$ become comparable because of rescattering.  We
are therefore forced to put the initial condition at the onset of
reionization.  As one can see, the temperature quadrupole produced by
reionization is fully connected with the polarization spectrum through the
initial condition $\Delta_2^T(k,t)$. This  forms the basis of our method : given
a model and the polarization autocorrelation spectrum $C_l^E$ one can get information
about the local quadrupole at reionization from which the quadrupole today
can be inferred.

One conceptual difficulty with the method is the following. Since the quadrupole
today $C_2^T(t_0)$ probes scales larger than the quadrupole at reionization
$C_2^T(t_r)$ it seems at first that there should not  be any way to make the
method work as,  strictly speaking, we are  measuring $C_2^T(t_r)$ which is
definitely not equal to $C_2^T(t_0)$. Examining the issue more carefully however,
one sees that there is significant overlap between the scales spanned by
 $\Delta_2^T(k,t_0)$ and $\Delta_2^T(k,t_r)$ as shown
 in Fig.~\ref{fig:quads}. Since the $l=2$ moment is mapped
into $k$-space by $j_2(x)$, we expect to get a wide range of scales
contributing to $C_2^T(t_0)$ as $j_2(x)$ is broadly distributed. Moreover
$\Delta_2^T(k,t_r)$ is also convolved with $G_l(k) =
\frac{3}{4}\sqrt{\frac{(l+2)!}{(l-2)!}} e^{\frac{3}{10}\tau_r} \int_{t_r}^{t_0}
dt \; \frac{j_l(x)}{x^2} \; \dot{\tau}e^{\frac{7}{10}\tau}$ as implied by
combining (\ref{eq:Emode}) and (\ref{eq:Polterm_sol}), which further increases
the overlap. The first five $G_l$'s as well as their multiplication with
$\Delta_2^T(k,t_r)$ are also shown in Fig.~\ref{fig:quads}.  Therefore even 
though what we
really measure is not $C_2^T$ but rather the contribution to $C_2^T$ coming
from the quadrupole at reionization, this is sufficient to reduce the error on
$C_2^T$ significantly below the cosmic variance limit.

One may also wonder why we could not  choose some very early time prior to
recombination to set the initial condition, as during tight coupling we have
$\Pi(k,t) = \frac{5}{2}\Delta_2^T(k,t)$~\cite{ZH} and so we can also relate the
quadrupole with in fact the whole of the polarization spectrum, not just the
part produced by reionization. The problem however in this case is that the
overlap between the local quadrupole at the early time and $\Delta_2^T(k,t_0)$
would be minuscule. The quadrupole today will be dominated by the particular
integral instead, and its error would be effectively cosmic variance again.

To get an estimate for the error on $C_2^T$ we let the amplitude of
$\Delta_2^T(k,t_r)$ vary as a free parameter. This can be modeled by
multiplying it with a parameter $q$ by hand, {\it i.e.} $\Delta_2^T(k,t_r)
\rightarrow q\Delta_2^T(k,t_r)$ in the two relevant equations
(\ref{eq:delta_2_sol}) and (\ref{eq:Polterm_sol}).  The variance $V_r$ of
$\Delta_2^T(k,t_r)$ can then be estimated as $V_r = V_q C_2^r$ where
$V_q = \text{Var}[q]$ can be obtained using the Fisher information matrix and 
where $C_2^r = \frac{1}{5}\sum_m \langle |\hat{a}_{2m}^r|^2 \rangle = 
\frac{2}{\pi}\int_0^\infty dk \; k^2 \;P_\Psi |\Delta_2^T(k,t_r)|^2$. Assuming
that the only error in the polarization spectrum comes from cosmic variance,
the Fisher matrix is a scalar given by

\begin{figure}
 \epsfig{file=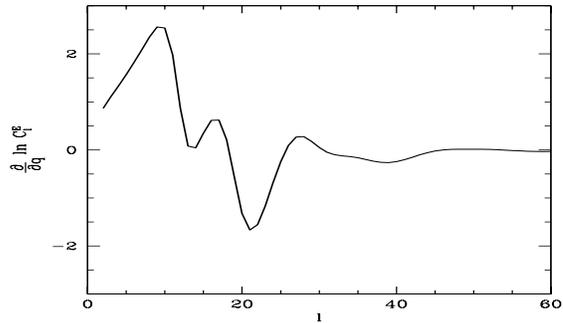,height=1.8in,width=3in}
\caption{The derivative of $\ln C_l^E$ with respect to $q$, the amplitude of
the local quadrupole at the onset of reionization.} 
\label{fig:dcl}
\end{figure}

\begin{equation}
   F = \sum_l (l + \frac{1}{2}) |\frac{\partial}{\partial q} \ln \;
   C_l^{E}|^2.
\end{equation}
The derivative in the expression above can be calculated numerically by
double-sided finite difference which we have taken to be $\delta q = 0.02$
around the fiducial value of $q=1$. This is shown in Fig.\ref{fig:dcl}.
The transfer functions were calculated
using a modified version of DASh~\cite{DASh}.  For the model considered we get 
 a variance of $q$ of $V_q= \frac{1}{F}= 0.002$.

Since we have used the Fisher matrix to get our estimate, we can assume that
the posterior pdf of $\hat{a}^r_{2m}$ is Gaussian, with variance $V_r$. Therefore
the posterior pdf of the total $\hat{a}_{2m} =
 e^{\frac{9}{10}\tau_r}\hat{a}_{2m}^r + \hat{a}_{2m}^{PI}$
will be Gaussian with variance $V_2 = e^{\frac{9}{5}\tau_r} V_r + C_2^{PI} + 2
e^{\frac{9}{10}\tau_r}C_2^{X}$, where
$C_2^{PI} = \frac{1}{5}\sum_m \langle |\hat{a}_{2m}^{PI}|^2 \rangle = 
\frac{2}{\pi}\int dk \; k^2 \; P_\Psi |\Delta_2^{PI}(k)|^2$ is the variance
of the particular integral and
$C_2^{X} = \frac{1}{10}\sum_m \langle \hat{a}_{2m}^{r}\hat{a}_{2m}^{* PI} + c.c. \rangle = 
\frac{2}{\pi} \int dk \; k^2 \; P_\Psi \Delta_2^T(k,t_r) \Delta_2^{PI}(k)$ the variance
of the correlation. Given that we have no extra information about the particular
integral we can assume that its variance will not change from the cosmic
variance value. The 
variance of $\hat{C}_2^T= \frac{1}{5}\sum_m |\hat{a}_{2m}|^2$ can then be estimated as 
$\text{Var}[\hat{C}_2^T] = \frac{2}{5}(V_2)^2$,
since under the above assumptions, $\hat{C}_2^T$ will obey a $\chi^2$
distribution with five degrees of freedom.  Our fiducial model gives
$\frac{\text{Var}[C_2^T]}{(C_2^T)^2} = 0.06$, which gives an error on $C_2^T$
of $24\%$.
   
One conceptual objection with the above argument is that in the usual case, one
has $\langle \hat{C}_2^T\rangle = C_2^T$ where as above it is equal to $V_2$ instead.
 This is not a problem as the true posterior pdf of
$\hat{C}_2^T$ would no longer be $\chi^2$ and therefore its mean and variance would
not be related in the usual way. 

\begin{figure}
 \epsfig{file=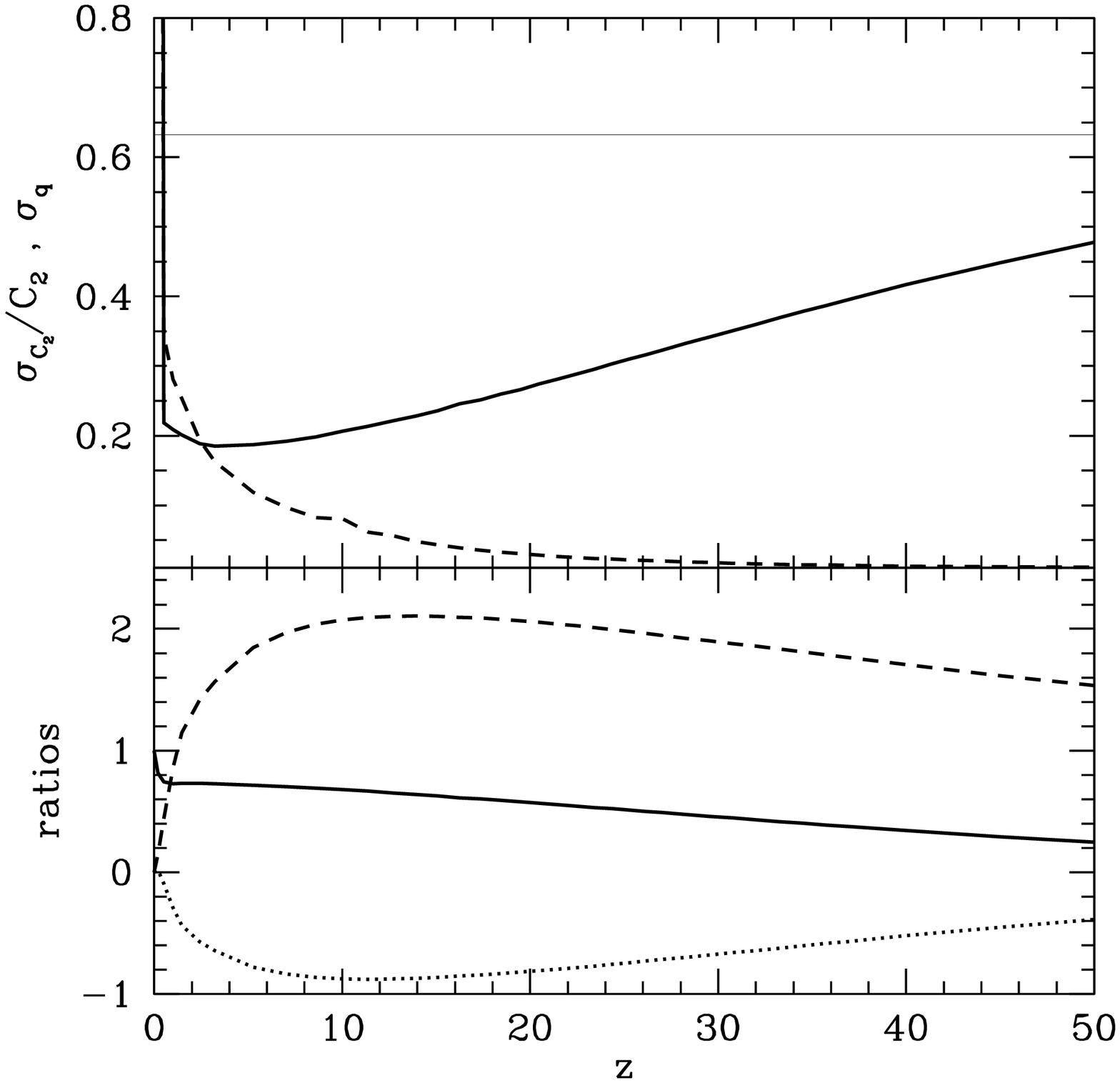,height=1.8in,width=3in}
\caption{Top panel : The variation of the quadrupole error 
$\frac{\sigma_{C_2^T}}{C_2^T}$(solid)  and $\sigma_q$ (dash) with
 reionization redshift. The error on the quadrupole reaches
a minimum as expected, around $z=3$. The cosmic variance limit is shown as a 
gray line. \protect{\newline}
 Bottom panel: Variation of the ratios  $\frac{C_2^r}{C_2^T}$ (solid) , 
$\frac{C_2^{PI}}{C_2^T}$(dash) and $\frac{C_2^X}{C_2^T}$(dotted) with reionization 
redshift.}
\label{fig:redshift}
\end{figure}

As argued before, the key point is the overlap of the $G_l(k)\Delta_2^T(k,t_r)$
with $\Delta_2^T(k,t_0)$. This signifies that the lower the reionization
redshift, the better the reduction of cosmic variance. Since however for low
reionization redshifts, we obtain  a smaller signal in the polarization spectrum, we would
also get a larger variance for the parameter $q$. We should therefore expect
that for low reionization redshifts, the error on $C_2^T$ would not be
improved but even be greater that cosmic variance. This means that as one
varies the reionization redshift $z_r$ from zero to some large value, the error on
$C_2^T$ would become better and better until some {\it conspiratory} value,
and then start to become worse and worse until it reaches cosmic variance
again. Another way to see this is the following. The correlation 
$\langle \hat{a}_{2m} \hat{a}_{2m}^r\rangle$ becomes arbitrarily small with increasing $z_r$,
as is implied from Fig.\ref{fig:quads} which means that we are probing many independent
Hubble volumes at that redshift hence the small variance of $q$. On the other hand  
$\frac{\langle \hat{C}_2 \hat{C}_2^r\rangle}{\langle (\hat{C}_2)^2\rangle} 
\approx \frac{C_2^r}{C_2^T}$ is also decreasing (but slowly enough)
 which means that the propagation of
our information on $C_2^r$ to $C_2^T$ becomes less effective at higher $z_r$~\footnote{
We thank Lloyd Knox for suggesting the calculation of these correlations.}.
 This is shown in Fig.(\ref{fig:redshift}).

The role of other cosmological parameters is also very important due to
imminent degeneracies, particularly with the optical depth. Still the quoted
error above is a lower bound and  other parameters can be
included at a later stage along with predictions for future polarization
experiments.

Finally let us  see how a different variance on $C_2^T$ could affect the
probabilities mentioned in the beginning. Strictly speaking, we need the true
posterior pdf of $C_2^T$ but based on our assumptions we can assume that it
would be approximately Gaussian with mean given by the model's value and
variance the value quoted above. Moreover, the maximum entropy principle,
would also give a Gaussian distribution if the only knowledge about the
distribution is the mean and the variance. If the fiducial model was the 
best fit model of a cosmic variance limited polarization experiment
then the probability that the quadrupole is as low as or lower than $250(\mu K)^2$, is
 reduced to $2\times10^{-4}$. 
This should be taken only as an order of magnitude
estimate of the actual probability which would have been given by the
true pdf. For comparison, using a Gaussian distribution with variance given by cosmic
variance instead, we get a probability which is only $8\%$ different than the
one quoted in the beginning.

We have shown that it is possible to reduce the error on the CMB temperature
quadrupole, to a value better than cosmic variance. The method exploits the
connection between the temperature quadrupole and the polarization spectrum
generated by a period of reionization.  This could reduce the variance of the
quadrupole significantly and has the potential to answer with confidence
whether the quadrupole is really low or not, compared to a given model. The
method is still at its infancy and further treatment is needed before
it can be incorporated with parameter estimation techniques. 
It would be an excellent way to test new physics.

\begin{acknowledgments}{\it Acknowledgments}
We thank Olivier Dor\'e, Joanna Dunkley, Pedro Ferreira and Lloyd Knox 
for very useful comments and penetrating questions.  CS is supported by a
Leverhulme foundation grant.
\end{acknowledgments}


\begin{thebibliography}{99}

\bibitem{WMAP} C.L. Bennett {\it et al.}, 2003, Astrophys. J. Supp. {\bf 148}, 1 (2003).
\bibitem{COBE} C.~L. Bennett {\it et al.}, Astrophys. J. Lett. {\bf 464}, L1 (1996)
; G.~Hinshaw {\it et al.}, Astrophys. J. Lett. {\bf 464}, L17 (1996).
\bibitem{WMAPfor} C.~L.~Bennett {\it et al.} Astrophys. J. Supp. {\bf 148}, 1
(2003).
\bibitem{BDFMS} M.~Bucher, J.~Dunkley, P.~G.~Ferreira, K.~Moodley and
C.~Skordis, astro-ph/0401417.
\bibitem{Efstathiou1} G. Efstathiou, Mon. Not. R. Astron. Soc. {\bf 346}, L26
(2003).
\bibitem{TdOCH} M.~Tegmark, A.~de~Oliveira-Costa and A.~Hamilton,
Phys. Rev. D {\bf 68}, 123523 (2003).
\bibitem{Efstathiou2} G. Efstathiou, astro-ph/0310207.
\bibitem{QML} M~.Tegmark, Phys. Rev. D {\bf 55}, 5895 (1997).
\bibitem{PCL} E.~Hivon {\it et al.}, Astrophys. J.{\bf 567}, 2 (2002); G.~Hinshaw
{\it et al.}, Astrophys. J. Supp. {\bf 148}, 135 (2003).
\bibitem{dOCTZH} A.~de~Oliveira-Costa, M.~Tegmark, M.~Zaldarriaga and
A.~Hamilton, astro-ph/0307282.
\bibitem{EHBGL} H.~K.~Eriksen {\it et al.}, astro-ph/0307507; 
 F.~K.~Hansen  {\it et al.},astro-ph/0402396
\bibitem{CHS} C.~J.~Copi, D.~Huterer and G.~D.~Starkman, astro-ph/0310511.
\bibitem{KL} M. Kamionkowski and A. Loeb, Phys.Rev. D {\bf 56}, 4511, (1997).
\bibitem{SZ} R.~A.~Sunyaev and Ya.~B.~Zel'dovich, Comments Astrophys. Space
Phys. {\bf 4}, 173 (1972); {\it ibid} Mon. Not. R. Astron. Soc. {\bf 190}, 413 (1980).
\bibitem{SS} N.~Seto and M.~Sasaki, Phys. Rev. D {\bf 62}, 123004 (2000).
\bibitem{CB} A.~Cooray and D.~Baumann, Phys.Rev. D {\bf 67}, 063505 (2003).
\bibitem{P} J. Portsmouth, astro-ph/0402173.
\bibitem{DHL} O. Dor\'e, G.~P. Holder and A. Loeb, astro-ph/0309281.
\bibitem{ZSS} M.~Zaldarriaga, D.~Spergel and U.~Seljak, Astrophys. J. {\bf 488}, 1
(1997).
\bibitem{reion} M.~Kaplinghat {\it et al.}, Astrophys.~J., {\bf 583}, 24 (2003);
G.~Holder {\it et al.}, Astrophys.~J. {\bf 595}, 13 (2003).
\bibitem{HH} W. Hu and G.~P. Holder, Phys.Rev. D {\bf 68}, 023001 (2003).
\bibitem{ZS} M.~Zaldarriaga and U.~Seljak, Phys.Rev. D {\bf 55}, 1830 (1997).
\bibitem{ZH} M.~Zaldarriaga and D.~Harari, Phys.Rev. D {\bf 52}, 3276 (1995).
\bibitem{DASh} M. Kaplinghat, L. Knox and C. Skordis, Astrophys.~J. {\bf 578}, 665
(2002).
\end{thebibliography}
\end{document}